\documentclass[a4paper,11pt]{article}
\usepackage{pos}
\usepackage{braket}
\usepackage{slashed}

\newcommand{\sm}{\text{SM}}
\newcommand{\np}{\text{NP}}
\newcommand{\lfu}{\text{LFU}}
\newcommand{\hqss}{\text{HQSS}}
\newcommand{\qcd}{\text{QCD}}
\newcommand{\ckm}{\text{CKM}}
\newcommand{\eff}{\text{eff}}
\newcommand{\hc}{\text{h.c.}}
\newcommand{\hami}{\mathcal{H}}
\newcommand{\oper}{\mathcal{O}}
\newcommand{\diff}{\text{d}}
\newcommand{\br}{\mathcal{B}}
\newcommand{\order}{\mathcal{O}}
\newcommand{\vslash}{\slashed{v}}
\newcommand{\tr}{\text{Tr}}
\newcommand{\kslash}{\slashed{k}}

\newcommand{\Drightslash}{\slashed{\vec{D}}}
\newcommand{\Dright}{\vec{D}}
\newcommand{\lagr}{\mathcal{L}}
\newcommand{\ff}{\text{FF}}
\newcommand{\nrqcd}{\text{NRQCD}}
\newcommand{\hq}{\text{HQ}}

\title{Investigating $B_c$ semileptonic decays}
\author*[a]{Francesco Loparco}
\affiliation[a]{Istituto Nazionale di Fisica Nucleare, Sezione di Bari,\\
Via Orabona 4, I-70126 Bari, Italy}

\emailAdd{francesco.loparco1@ba.infn.it}
\abstract{
I present analyses of semileptonic $B_c$ decays.
I first discuss the decays induced by $c \to \set{s, d}$ transitions, in particular $B_c \to B_a \, \bar{\ell} \, \nu_\ell$ and $B_c \to B_a^*(\to B_a \, \gamma) \, \bar{\ell} \, \nu_\ell$ decays, with $a = \set{s, d}$ and $\ell = \set{e, \mu}$, in the Standard Model and in the extension based on the low-energy Hamiltonian comprising the full set of $D = 6$ semileptonic $c \to s, d$ operators with left-handed neutrinos.\\
Moreover, I consider $b \to c$ modes, with particular focus on the determination of  the form factors parametrizing the $B_c \to J/\psi,\eta_c$ matrix elements of the operators in a generalized low-energy $b \to c$ semileptonic Hamiltonian.
In this case I consider an expansion in nonrelativistic \qcd\ together with an expansion in the inverse heavy-quark mass which allows the form factors to be expressed in terms of universal functions in a selected kinematical range.\\
The Heavy Quark Spin Symmetry is used for both analyses.
In the first case,  it allows the relevant hadronic matrix elements to be related and the lattice \qcd\ results on $B_c$ form factors to be exploited.
Optimized observables are selected, and correlations among them is studied to identify the effects of the various operators in the extended low-energy Hamiltonian.
In the second one, using as an input the lattice \qcd\ results for the $B_c \to J/\psi$ matrix element of the Standard Model operator, it is possible to obtain information on other form factors.
The extrapolation to the full kinematical range is also presented.
}

\FullConference{%
  41st International Conference on High Energy physics - ICHEP2022\\
  6-13 July, 2022\\
  Bologna, Italy
}

\begin{document}
\maketitle

\section{Introduction}

The $B_c$ meson was observed for the first time by the CDF Collaboration \cite{Abe:1998ihx}.
%It has the structure of a heavy quarkonium but decays weakly.
%So that it is well suited to study both quarkonium and weak interaction features.
It can decay through the charm transitions $c \to s,d$ that dominate over the $b \to c,u$ decays and over the annihilation mode.
This hierarchy is due to of the large values the moduli of the \ckm\ matrix elements $|V_{cs}|$ and $|V_{cd}|$, despite the smaller available phase-space \cite{Colangelo:1992cx, Beneke:1996xe, Anisimov:1998uk, Kiselev:2000pp}.
Among transitions due to charm quark decays, I present a study of the exclusive semileptonic modes $B_c \to B_{s, d}^{(*)} \, \bar{\ell} \, \nu_\ell$ with $\ell = \set{e, \mu}$ (the $\tau$ mode is phase-space forbidden), that can be useful to investigate $\mu / e$ universality.

The study of the beauty transitions for $B_c$ represents another way to investigate flavour anomalies detected in several $b \to c$ decays \cite{Amhis:2019otj, Gambino:2020jvv}.
If such deviations from the Standard Model (\sm) are due to New Physics (\np) phenomena violating lepton flavour universality (\lfu), analogous effects should be found in $B_s$, $B_c$ and $b$-baryon decay modes, both inclusive and exclusive \cite{Colangelo:2016ymy, Colangelo:2020vhu}.
However, because of the different hadronic uncertainties affecting the various processes, it is necessary to analyze each mode separately.
I discuss the exclusive semileptonic $B_c \to J/\psi(\eta_c) \, \ell \, \bar{\nu}_\ell$ decays which are under experimental scrutiny \cite{Aaij:2017tyk}.

To study these processes one can invoke the Heavy Quark Spin Symmetry (\hqss) \cite{Jenkins:1992nb}, which allows, in the case of $B_c \to B_{s, d}^{(*)}$ modes, all the form factors to be expressed in terms of two independent functions.
Such functions can be derived from the $B_c \to B_{s, d}$ form factors, determined by lattice \qcd\ \cite{Cooper:2020wnj} and can be used those parametrizing $B_c \to B_{s, d}^*$ modes.
Moreover, exploiting the methods of non relativistic \qcd\ (\nrqcd), relations can be derived among the $B_c \to J/\psi$ 
%\cite{Harrison:2020gvo}
and $B_c \to \eta_c$ form factors, both in the case of the matrix elements of the \sm\ operators, and when \np\ operators are considered, as in the generalized Hamiltonians considered below.

\vspace{-0.2cm}

\section{Exclusive $c \to \set{s, d} \, \bar{\ell} \, \nu_\ell$ modes: $B_c^+ \to B_{s, d}^{(*)} (\to B_{s, d} \, \gamma) \, \bar{\ell} \, \nu_\ell$}

We consider the low-energy Hamiltonian comprising the full set of $D = 6$ semileptonic $c \to a = \set{s, d}$ operators with left-handed neutrinos
\begin{align}
\label{hamiltonian_c_to_D}
\hami_\eff^{c \to a \, \bar{\ell} \, \nu_\ell} = \frac{G_F \, V_{ca}^*}{\sqrt{2}} \, \left[ ( 1 + \epsilon_V^\ell ) \, \oper_\sm + \epsilon_S^\ell \, \oper_S + \epsilon_P^\ell \, \oper_P + \epsilon_T^\ell \, \oper_T + \epsilon_R^\ell \, \oper_R \right] + \hc \;.
\end{align}
$G_F$ is the Fermi constant and $V_{ca}$ is the relevant \ckm\ matrix element.
In addition to the \sm\ operator $\oper_\sm = \big[ \bar{a} \, ( 1 + \gamma_5 ) \, \gamma_\mu \, c \big] \, \big[ \bar{\nu}_\ell \, ( 1 + \gamma_5 ) \, \gamma^\mu \, \ell \big]$ we consider: $\oper_S = \big[ \bar{a} \, c \big] \, \big[ \bar{\nu}_{\ell} \, ( 1 + \gamma_5 ) \, \ell \big]$, $\oper_P = \big[ \bar{a} \, \gamma_5 \, c \big] \, \big[ \bar{\nu}_{\ell} \, ( 1 + \gamma_5 ) \, \ell \big]$, $\oper_T = \big[ \bar{a} \, ( 1 + \gamma_5 ) \, \sigma_{\mu\nu} \, c \big] \, \big[ \bar{\nu}_{\ell} \, ( 1 + \gamma_5 ) \, \sigma^{\mu\nu} \, \ell \big]$ and $\oper_R = \big[ \bar{a} \, ( 1 - \gamma_5 ) \, \gamma_\mu \, c \big] \, \big[ \bar{\nu}_\ell \, ( 1 + \gamma_5 ) \, \gamma^\mu \, \ell \big]$.
$\epsilon_{V, S, P, T, R}^\ell$ are complex lepton-flavour dependent couplings.
In the case of $\epsilon_i^\ell = 0$ one recovers the \sm\ case.
We keep $m_\ell \neq 0$ for $\ell = \set{e, \mu}$.

The $q^2$ distribution of the $B_c \to P \, \bar{\ell} \, \nu_\ell$ decay, with $P$ a pseudoscalar meson, governed by the low-energy Hamiltonian \eqref{hamiltonian_c_to_D} reads:
\begin{align}
& \frac{\diff \Gamma (B_c \to P \, \bar{\ell} \, \nu_\ell)}{\diff q^2} = \frac{G_F^2 \, |V_{ca}|^2 \, \sqrt{\lambda}}{128 \, m_{B_c}^3 \, \pi^3 \, q^2} \, \bigg( 1 - \frac{m_\ell^2}{q^2} \bigg)^2 \, \Big\{ \Big| m_\ell \, ( 1 + \epsilon_V^\ell + \epsilon_R^\ell ) + \epsilon_S^\ell \, \frac{q^2}{m_c - m_a} \Big|^2 \, ( m_{B_c}^2 - m_P^2 ) \, f_0^2 + \notag \\
& \quad + \lambda \, \Big[ \frac{1}{3} \, \Big| m_\ell \, ( 1 + \epsilon_V^\ell + \epsilon_R^\ell ) \, f_+ + \epsilon_T^\ell \, \frac{4 \, q^2}{m_{B_c} + m_P} \, f_T \Big|^2 + \frac{2 \, q^2}{3} \, \Big| ( 1 + \epsilon_V^\ell + \epsilon_R^\ell ) \, f_+ + \epsilon_T^\ell \, \frac{4 \, q^2}{m_{B_c} + m_P} \, f_T \Big|^2 \Big] \Big\} \;,
\end{align}
where $q^2$ is the squared momentum transferred to the lepton pair and $\lambda = \lambda(m_{B_c}^2, m_P^2, q^2)$ is the triangular function.
The definition of the form factors (\ff) $f_{+, 0, T} = f_{+, 0, T}(q^2)$ can be found in \cite{Colangelo:2021dnv}.

\begin{figure}[h]
\centering
\includegraphics[width=6cm,clip]{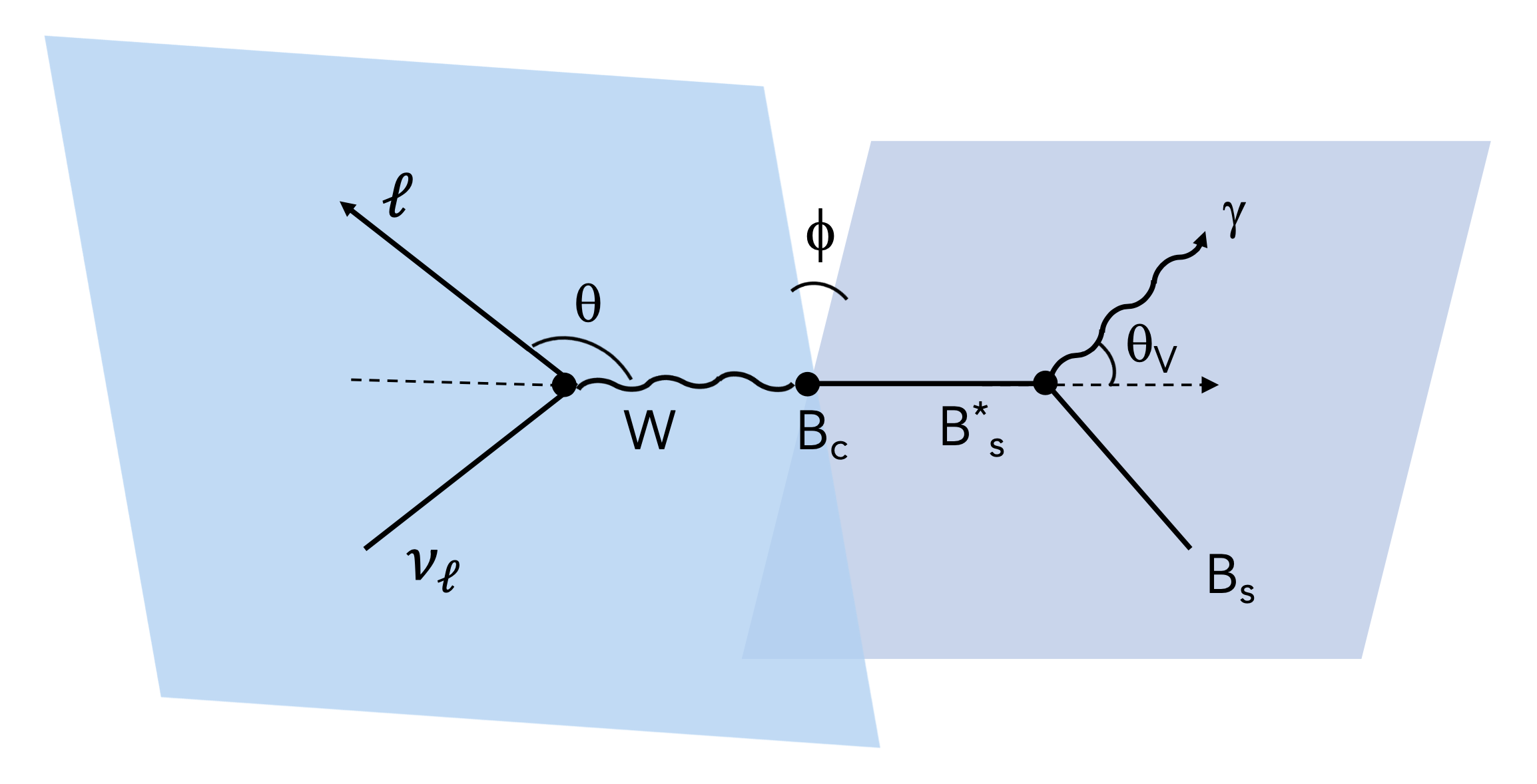}
\caption{Kinematics of the $B_c \to B_s^* ( \to B_s \, \gamma ) \, \bar{\ell} \, \nu_\ell$ decay.}
\label{fig1}
\end{figure}

In the case of $B_c \to V ( \to P \, \gamma ) \, \bar{\ell} \, \nu_\ell$, the four-body kinematics is shown in figure \ref{fig1}.
The fully differential decay width, obtained in the narrow width approximation for the meson $V$, reads:
\begin{align}
\label{distr}
& \frac{\diff^4 \Gamma(B_c \to V(\to P \, \gamma) \, \bar{\ell} \, \nu_\ell)}{\diff q^2 \, \diff \cos\theta \, \diff \phi \, \diff \cos\theta_V} = \mathcal{N} \, | \vec{p}_V | \, \left( 1 - \frac{m_\ell^2}{q^2} \right)^2 \times \big\{ I_{1s} \, \sin^2 \theta_V + I_{1c} \, (3 + \cos 2 \theta_V) + \notag \\
& \quad + (I_{2s} \, \sin^2 \theta_V + I_{2c} \, (3 + \cos 2 \theta_V)) \, \cos 2\theta + I_{3} \, \sin^2 \theta_V \, sin^2 \theta \, \cos 2\phi + I_{4} \, \sin 2\theta_V \, sin 2\theta \, \cos \phi + \notag \\
& \quad + I_{5} \, \sin 2\theta_V \, sin \theta \, \cos \phi + (I_{6s} \, \sin^2 \theta_V + I_{6c} \, (3 + \cos 2 \theta_V)) \, \cos \theta + \notag \\
& \quad + I_{7} \, \sin 2\theta_V \, sin \theta \, \sin \phi + I_{8} \, \sin 2\theta_V \, sin 2\theta \, \sin \phi + I_{9} \, \sin^2 \theta_V \, sin^2 \theta \, \sin 2\phi \big\} \;,
\end{align}
with $\mathcal{N} = \frac{3 \,  G_F^2 \, |V_{ca}|^2 \, \br(V \to P \, \gamma)}{128 \, (2 \, \pi)^4 \, m_{B_c}^2}$ and $| \vec{p}_V | = \sqrt{\lambda(m_{B_c}^2, m_V^2, q^2)} / 2 \, m_{B_c}$.
The functions $I_i = I_i(q^2)$ encode the dynamics of the \sm\ and of \np\ described in \eqref{hamiltonian_c_to_D} and are derived in \cite{Colangelo:2021dnv}.

In the heavy-quark mass limit $m_Q \gg \Lambda_\qcd$ the \qcd\ Lagrangian exhibits the \hqss\ \cite{Neubert:1993mb} which produces the decoupling of the spins of the heavy quarks in $B_c$: spin-spin interaction vanishes in this limit.
Consequently, relations among the \ff\ parametrizing the weak-current matrix elements can be obtained.
In the semileptonic $B_c \to B_a^{(*)}$ decays, since $m_c \ll m_b$, the energy released to the final hadronic system is much smaller than $m_b$, so that the final meson keeps the same $B_c$ four-velocity.
Denoting the initial- and final-meson four-momenta as $p = m_{B_c} \, v$ and $p' = m_{B_a^{(*)}} \, v' = m_{B_a^{(*)}} \, v+k$, with $k$ a small residual momentum, the four-momentum transferred to the leptons is $q = p - p' = ( m_{B_c} - m_{B_a^{(*)}} ) \, v - k$, with $v \cdot k = \order( 1 / m_b )$.
The heavy pseudoscalar and vector mesons are collected in doublets, the two components of which represent states differing only for the orientation of the
heavy quark spins \cite{Falk:1991nq}:
\begin{align}
\label{operator_H}
H^{c \bar{b}} = P_+(v) \, \big[ B_c^{*\mu} \, \gamma_\mu - B_c \, \gamma_5 \big] \, P_-(v) \hspace{0.5cm} \text{and} \hspace{0.5cm} H_a^{\bar{b}} = \big[ B_a^{*\mu} \, \gamma_\mu - B_a \, \gamma_5 \big] \, P_-(v) \;,
\end{align}
where $P_\pm(v) = \frac{1 \pm \vslash}{2}$.
The two doublets correspond to $( B_c^+, B_c^{*+} )$ and $( B_a, B_a^* )$, respectively.

Exploiting the trace formalism \cite{Falk:1990yz}, the matrix elements of the quark current $\bar{a} \, \Gamma \, c$ between $B_c$ and $B_a^{(*)}$, with $\Gamma$ a generic product of Dirac matrices, can be written as
\begin{align}
\label{trace}
\braket{B_a^{(*)}(v, k, (\epsilon)) | \bar{a} \, \Gamma \, c | B_c(v)} = - \sqrt{m_{B_c} \, m_{B_a^{(*)}}} \, \tr \big[ \overline{H}_a^{\bar{b}} \, \Omega_a(v, a_0 \, k) \, \Gamma \, H^{c \bar{b}} \big] \;,
\end{align}
%In \eqref{trace}, $\overline{H}_a^{\bar{b}} = \gamma^0 \, H_a^{\bar{b}} \, \gamma^0$ and $H^{c \bar{b}}$ are invariant under rotations of the $\bar{b}$ spin.
where $\Omega_a(v, a_0 \, k) = \Omega_{1a} + \kslash \, a_0 \, \Omega_{2a}$ involves two dimensionless functions, $\Omega_{1a}$ and $\Omega_{2a}$.
The dimensionful parameter $a_0$ can be identified with the length scale of the process, e.g. the Bohr radius of the mesons.
The general parametrization of the matrix elements of the operators in \eqref{hamiltonian_c_to_D} involves several \ff.
Using this formalism, one can write them in terms of $\Omega_{1a}$ and $\Omega_{2a}$, reducing the number of independent hadronic functions \cite{Colangelo:2021dnv}.
Spin symmetry relations strictly hold close to the zero-recoil point where the produced meson is at rest in the $B_c$ rest frame \cite{Colangelo:1999zn}.
However, the relatively small phase space justifies the extrapolation to the full kinematical range.

\begin{figure}[b]
\centering
\includegraphics[width=6cm,clip]{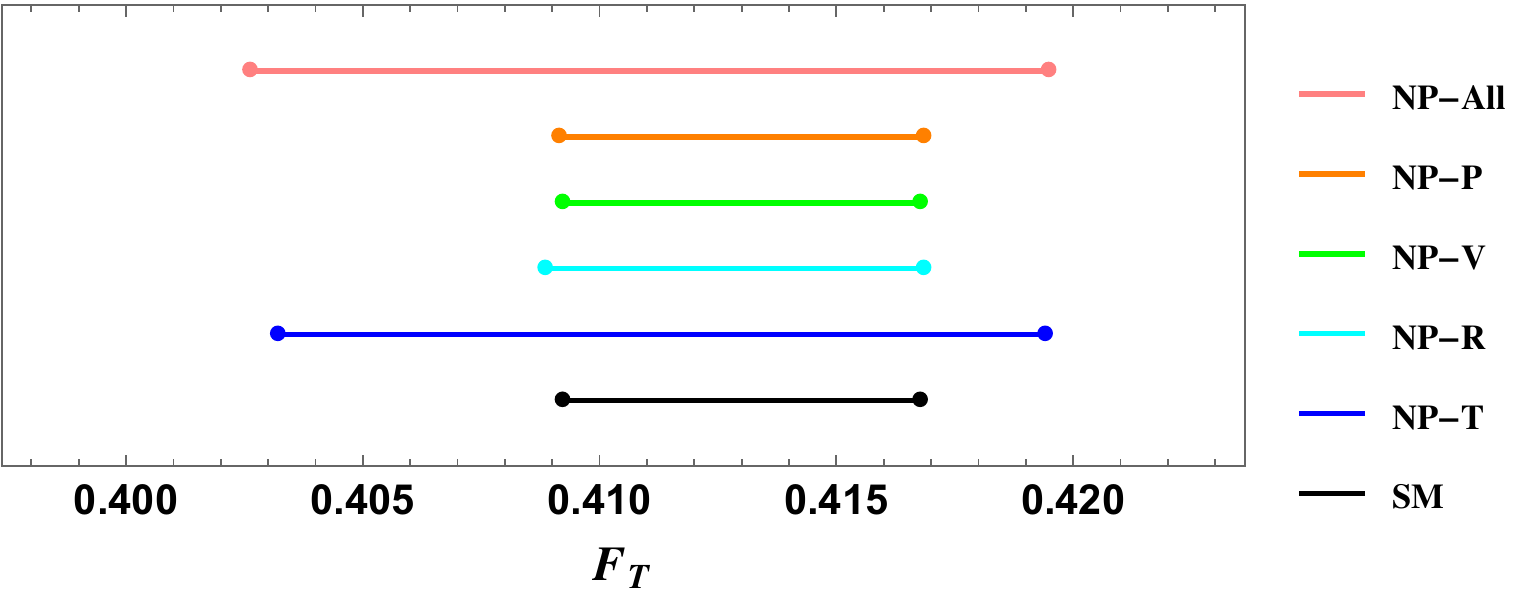}
\hspace{0.5cm}
\includegraphics[width=6cm,clip]{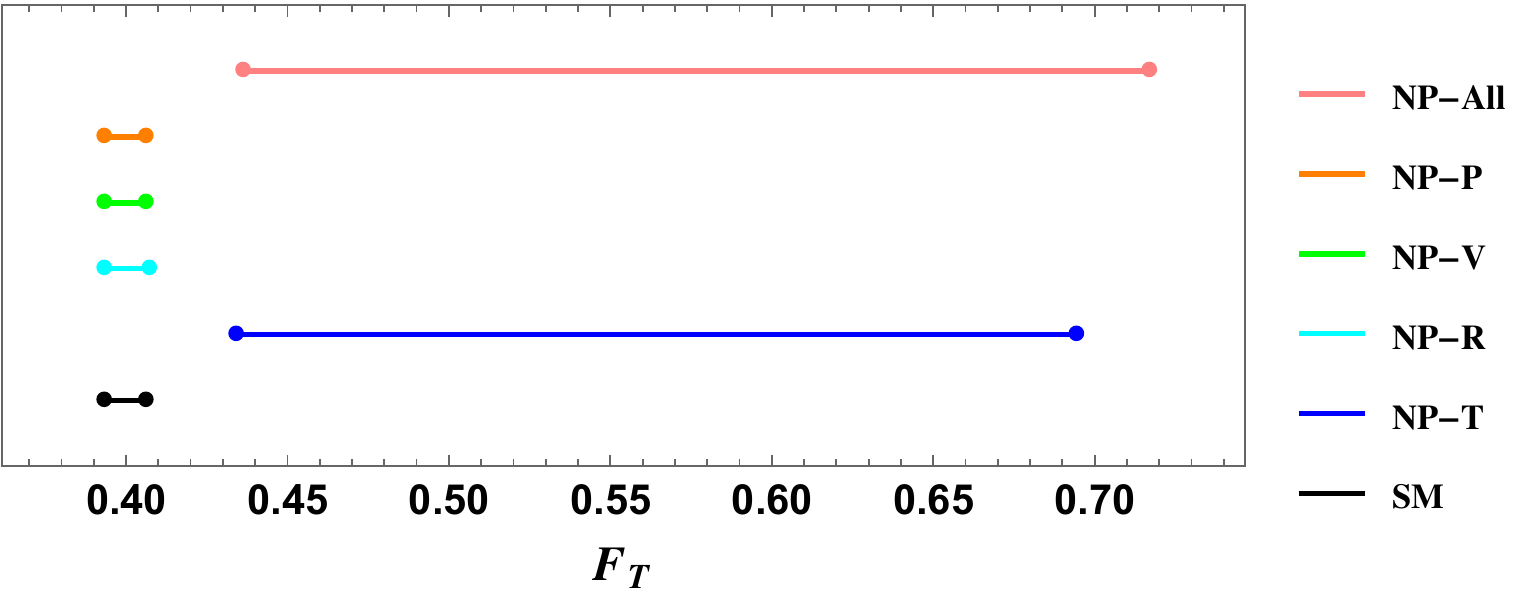}
\caption{$F_T$ for $B_s^*$ (left) and $B_d^*$ (right) in the \sm\ and including of \np\ operators.}
\label{fig2}
\end{figure}

In our numerical analysis of $B_c \to B_{s,d}^{(*)}$ we determine the universal functions $\Omega_{1s(d)}$ and $\Omega_{2s(d)}$ \cite{Colangelo:2021dnv} using the lattice \qcd\ results for the \ff\ $f_{+, 0}^{B_c \to B_s}$ and $f_{+, 0}^{B_c \to B_d}$ \cite{Cooper:2020wnj}.\\
Using such results we obtain the branching fractions in \sm\ (table \ref{tab1}) \cite{Colangelo:2021dnv}.
\begin{table}[t]
\centering
\caption{Branching fractions of $B_c \to B_{s, d}^{(*)} \, \bar{\ell} \, \nu_\ell$ in \sm.
$x_s = |V_{cs} / 0.987|^2$ and $x_d = | V_{cd} / 0.221 |^2$.}
\small
\label{tab1}
\begin{tabular}{ccc}
\hline
$(a, \ell)$ & $\br(B_c^+ \to B_a \, \ell^+ \, \nu_\ell)$ & $\br(B_c^+ \to B_a^* \, \ell^+ \, \nu_\ell)$ \\
\hline
$(s, \mu)$ & $1.25(4) \times 10^{-2} \, x_s$ & $3.0(1) \times 10^{-2} \, x_s$ \\
$(s, e)$ & $1.31(4) \times 10^{-2} \, x_s$ & $3.2(1) \times 10^{-2} \, x_s$ \\
$(d, \mu)$ & $8.3(5) \times 10^{-4} \, x_d$ & $20(1) \times 10^{-4} \, x_d$ \\
$(d, e)$ & $8.7(5) \times 10^{-4} \, x_d$ & $21(1) \times 10^{-4} \, x_d$ \\
\hline
\end{tabular}
\end{table}
Another interesting observable is the ratio $F_T = \frac{\Gamma_T}{\Gamma_T + \Gamma_L}$, where $\Gamma_{T, L}$ are the decay widths to transversely and longitudinally polarized $B_{s, d}^*$, respectively.
In figure \ref{fig2} we display our results for $F_T$ in \sm\ and \np\ using the ranges fixed in \cite{Becirevic:2020rzi} for the couplings $\epsilon_i^\ell$ in \eqref{hamiltonian_c_to_D}.

\vspace{-0.2cm}

\section{Exclusive $b \to c \, \ell \, \bar{\nu}_\ell$ modes: $B_c^- \to J/\psi (\eta_c) \, \ell \, \bar{\nu}_\ell$}

In order to study $B_c \to J/\psi, \eta_c$ semileptonic decays we extend the previous analysis at the next-to-leading order (NLO) in the expansion, exploiting in addition to the \hqss\ the power counting rules of \nrqcd. 
In this way we express the \ff\ in terms of a set of universal functions.
These relations  can be used to test the \ff\ obtained by different methods.

To construct the \hq\ expansion, the \qcd\ field $Q(x)$ with mass $m_Q$ is written:
\begin{align}
\label{decom}
Q(x) = e^{i \, m_Q \, v \cdot x} \, \psi(x) = e^{i \, m_Q \, v \cdot x} \, \big( \psi_+(x) + \psi_-(x) \big) \;,
\end{align}
where $\psi_\pm(x) = P_\pm(v) \, \psi(x)$; $\psi_+$ is the positive energy component of the field \cite{Aebischer:2021ilm}; $v$ is the heavy-meson (quarkonium four-velocity) and
%the \eom\
%\begin{align}
%\label{eom}
$\psi_-(x) = \frac{1}{2 \, m_Q + i \, v \cdot \Dright} \, i \, \Drightslash_\perp \, \psi_+(x)$,
%\end{align}
with $\Dright_{\perp\mu} = D_\mu - ( v \cdot D ) \, v_\mu$.
The various operators can be ordered within NRQCD according to their scaling with $\tilde{v}$, the relative \hq\ three-velocity in the hadron rest frame, satisfying the relation $\tilde{v} = | \vec{\tilde{v}} | \ll 1$ \cite{Bodwin:1994jh}.
Also the \qcd\ Lagrangian density can be expanded and divided into a leading order (LO) term and a NLO correction
\begin{align}
\lagr_\qcd = \bar{\psi}_+ \, \bigg( i \, v \cdot \Dright + \frac{( i \, \Dright_\perp )^2}{2 \, m_Q} + \frac{g \, \sigma \cdot G_\perp}{4 \, m_Q} + \frac{i \, \Drightslash \, ( - i \, v \cdot \Dright ) \, i \, \Drightslash}{4 \, m_Q^2} + \dots \bigg) \, \psi_+ = \lagr_0 + \lagr_1 + \dots \;.
\end{align}
The detailed calculation can be found in \cite{Colangelo:2022lpy}.
The expansion of a generic weak current $\bar{Q}' \, \Gamma \, Q$ is:
\begin{align}
\label{current}
\bar{Q}' \, \Gamma \, Q = J_0 + \bigg( \frac{J_{0,1}}{2 \, m_Q} + \frac{J_{1,0}}{2 \, m_{Q'}} \bigg) + \bigg( - \frac{J_{0,2}}{4 \, m_Q^2} - \frac{J_{2,0}}{4 \, m_{Q'}^2} + \frac{J_{1,1}}{4 \, m_Q \, m_{Q'}} \bigg) + \order(1 / m^3) \;,
\end{align}
where $m$ denotes the masses of the heavy quarks and the currents $J_i$ defined in \cite{Colangelo:2022lpy}.

The $B_c \to J/\psi$, $\eta_c$ matrix elements of the various currents in \eqref{current} can be expressed using the trace formalism \cite{Falk:1991nq}, describing the lowest-lying S-wave $\bar{b} c$ and $\bar{c} c$ bound states as in \eqref{operator_H} \cite{Jenkins:1992nb}.

\begin{figure}[h]
\centering
\includegraphics[width=4cm,clip]{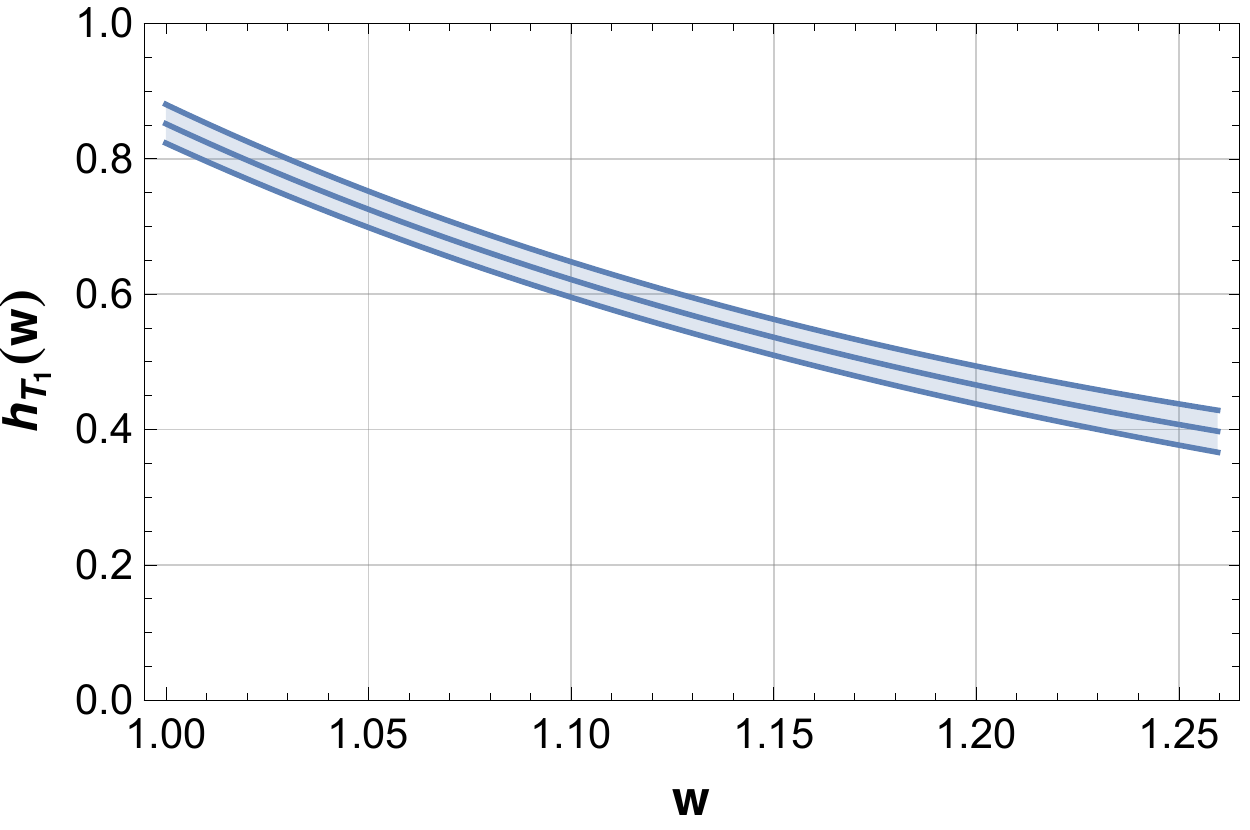}
\hspace{0.5cm}
\includegraphics[width=4cm,clip]{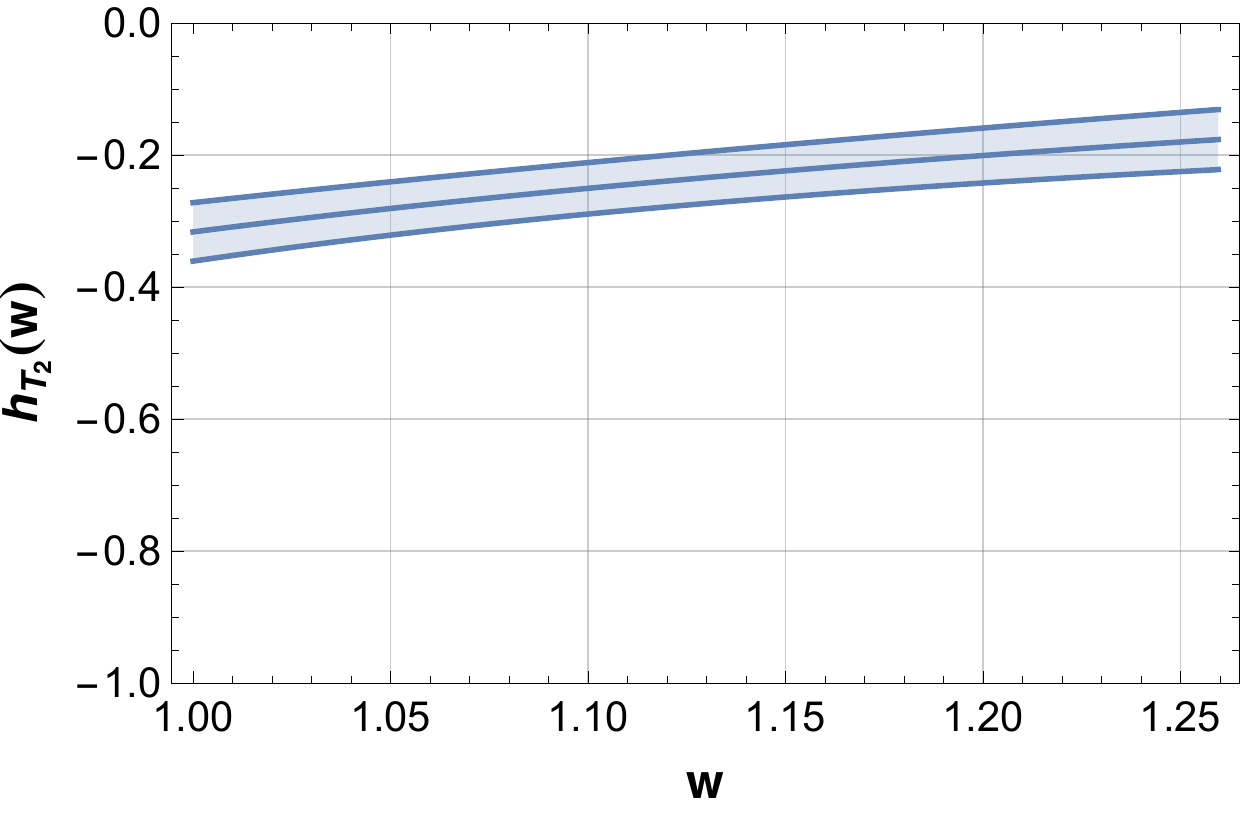}
\hspace{3cm}
\includegraphics[width=4cm,clip]{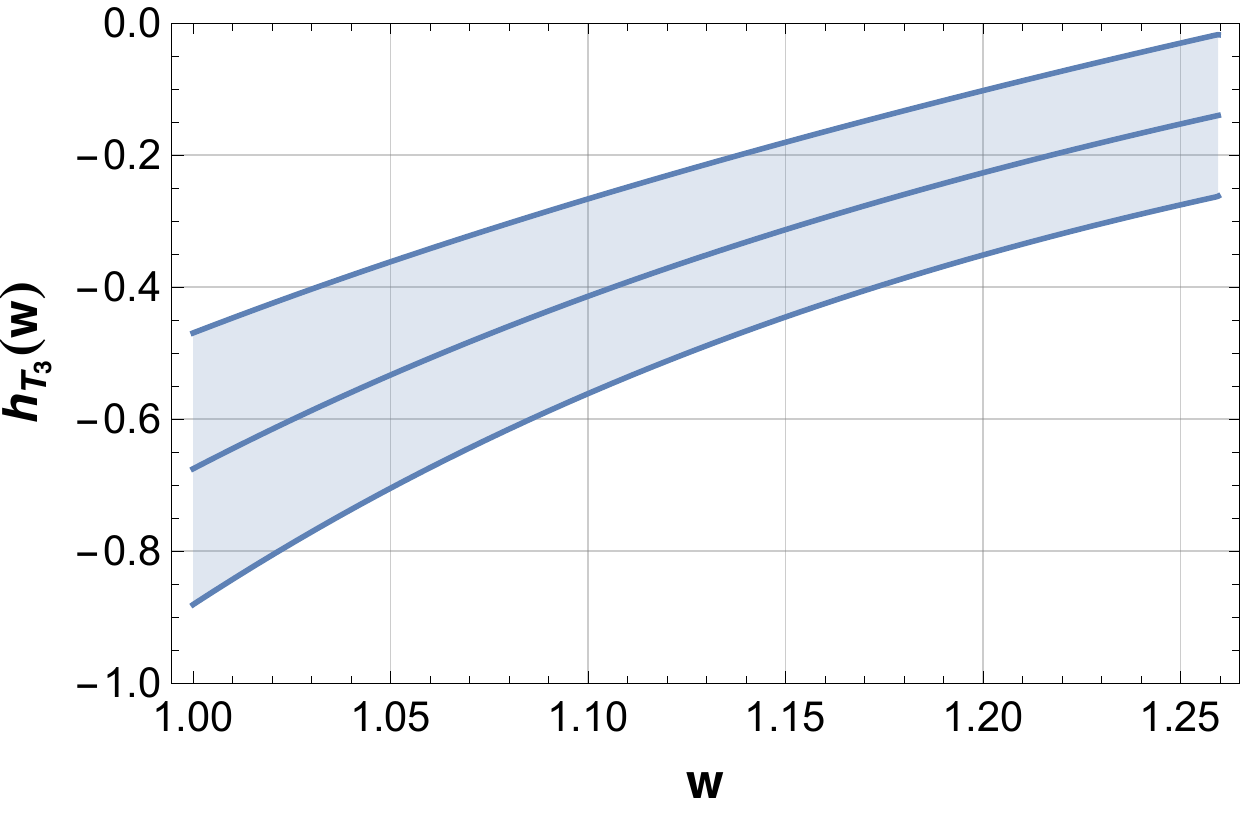}
\hspace{0.5cm}
\includegraphics[width=4cm,clip]{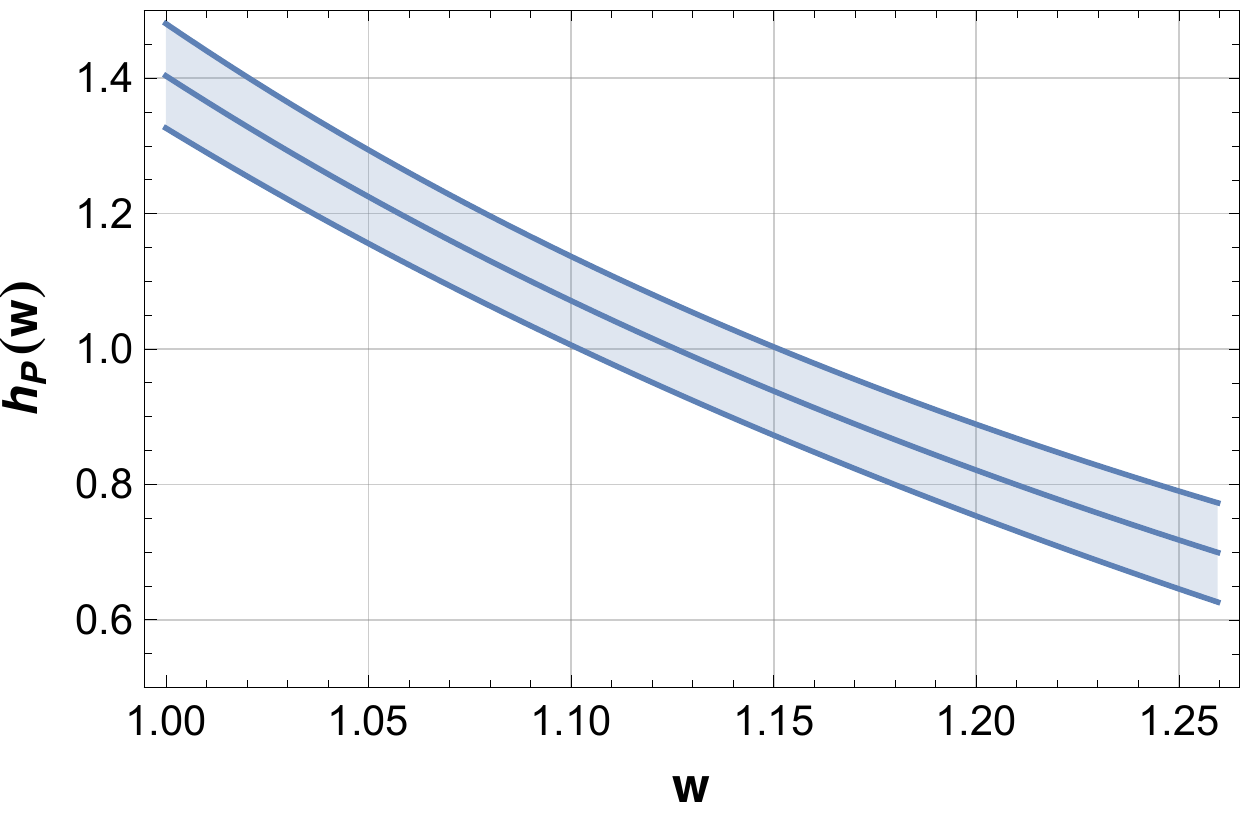}
\hspace{5cm}
%\caption{$B_c \to J/\psi$ form factors.}
%\label{fig3}
%\end{figure}
%\begin{figure}[h!]
%\centering
\includegraphics[width=4cm,clip]{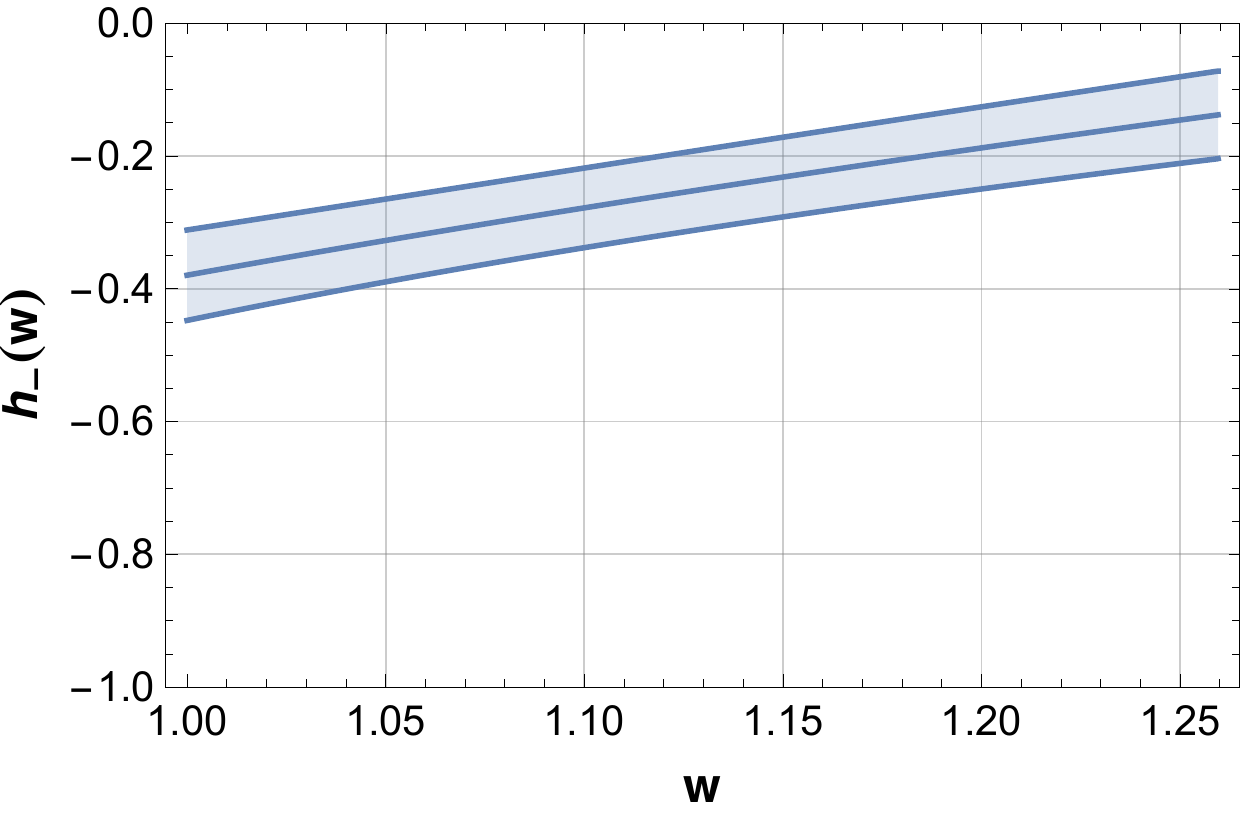}
\hspace{0.5cm}
\includegraphics[width=4cm,clip]{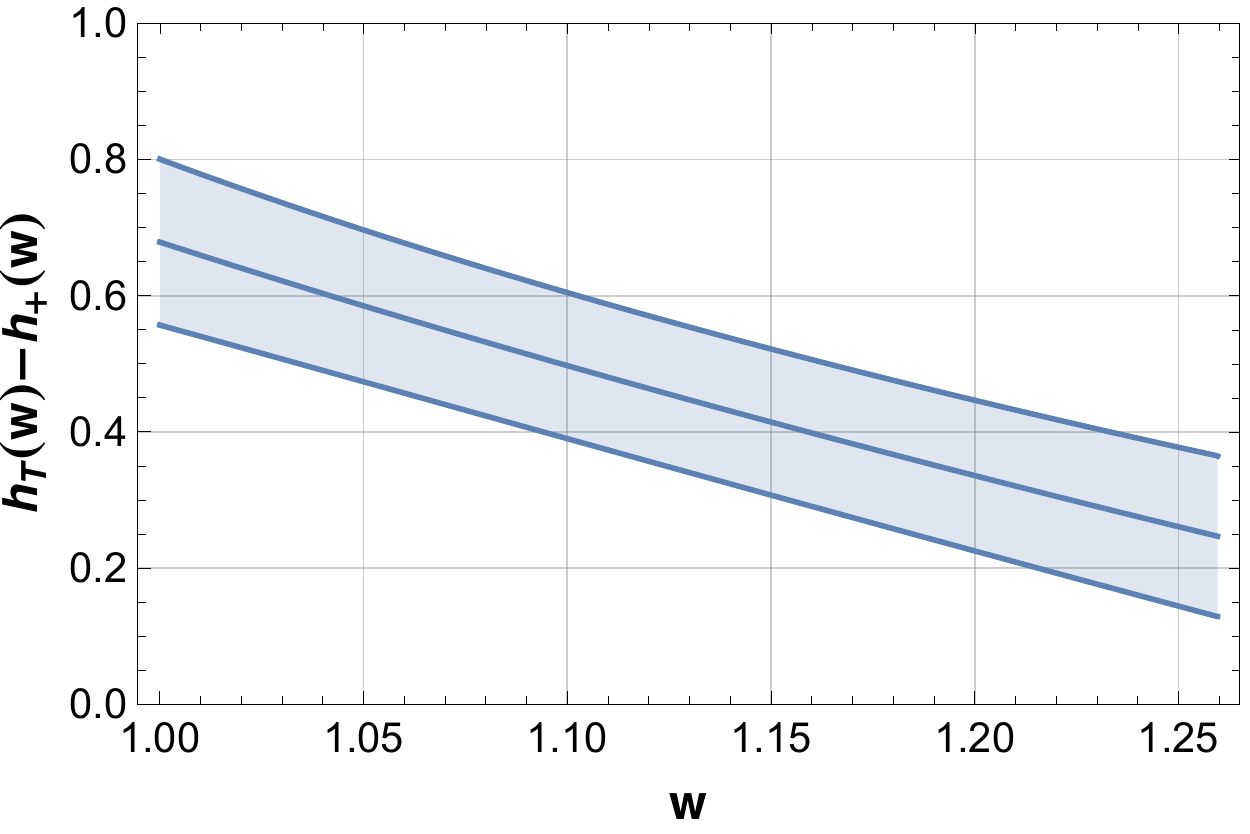}
\hspace{0.5cm}
\includegraphics[width=4cm,clip]{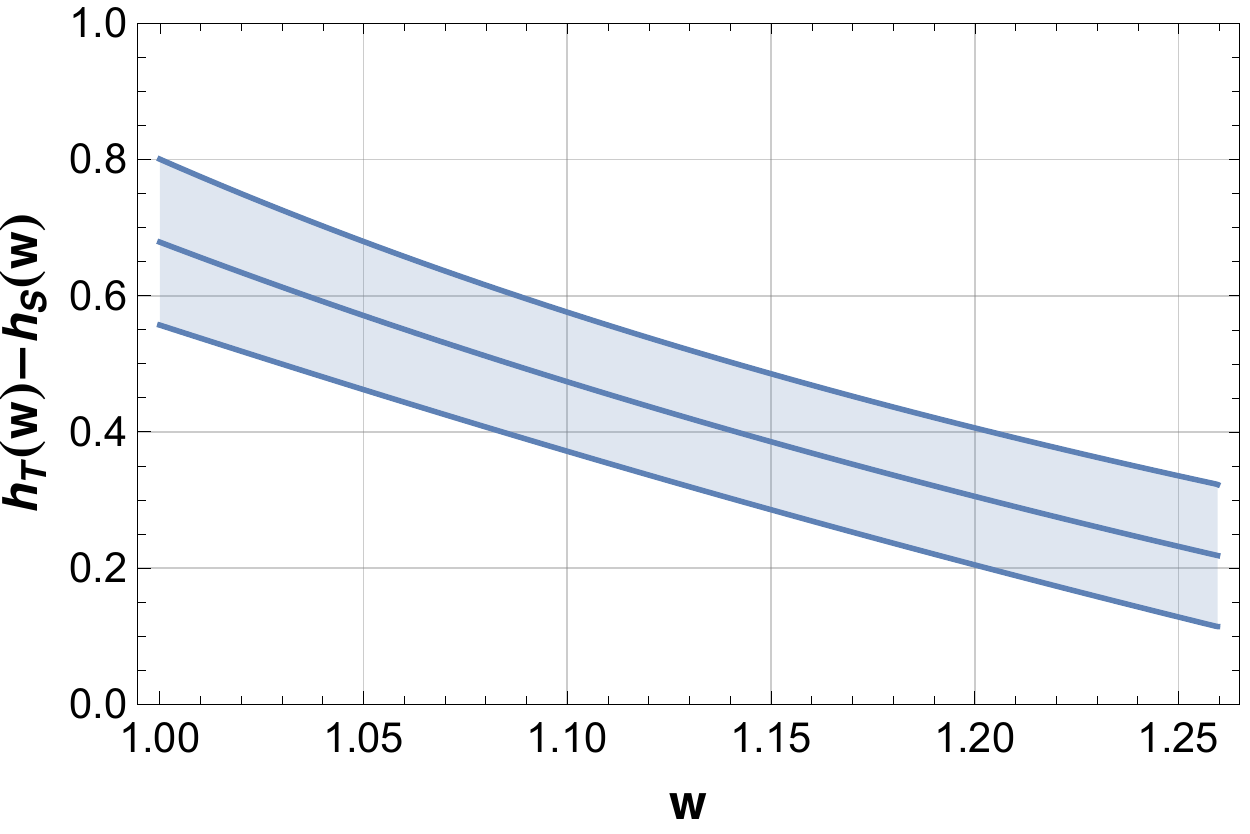}
\caption{$B_c \to J/\psi$ form factors (upper and middle plots) and $B_c \to \eta_c$ form factors (lower plots).}
\label{fig3}
\end{figure}

Going beyond LO the number of universal functions increases.
In our numerical analysis we have included terms  up to $\order(\tilde{v}^3)$ and to $\order(1/m)$.
Exploiting the lattice results for $V(q^2)$ and $A_{1,2,0}(q^2)$ \cite{Harrison:2020gvo} and the relations among the $B_c \to J/\psi$ and $B_c \to \eta_c$ \ff\ derived using the formalism described above, we obtain the \ff\ displayed in figure \ref{fig3} as a function of $w = v \cdot v'$.

\vspace{-0.4cm}

\section{Conclusions}

Semileptonic $B_c$ decays play an interesting role in the \sm\ and in the investigation of flavour anomalies as those observed in other beauty hadron decays.
Exploiting an effective approach based on the \hqss\ and \nrqcd\ methods it is possible to find relations among the hadronic form factors, reducing the related uncertainties.
The same method can be applied to describe $B_c$ to P-wave charmonia \cite{Colangelo:2022awx}.

\vspace{-0.2cm}

\section*{Acknowledgements}

%\textbf{Acknowledgements}
I thank P. Colangelo, F. De Fazio, N. Losacco, M. Novoa-Brunet for collaboration.
This study has been  carried out within the INFN project QFT-HEP.

\nocite{}

\bibliographystyle{JHEP}
\bibliography{biblio}

\end{document}